\documentclass[
aps,
pra,
reprint,
groupeaddress]{revtex4-1}
\usepackage{graphicx}
\usepackage{color}
\usepackage{bbm}
\usepackage{amsfonts}
\usepackage{amssymb}
\usepackage{amsmath}
\newcommand{\be}[1]{\begin{equation}\label{#1}}
\newcommand{\ee}{\end{equation}}
\newcommand{\bea}[1]{\begin{eqnarray}\label{#1}}
\newcommand{\eea}{\end{eqnarray}}

\newcommand{\Eq}[1]{Eq.(\ref{#1})}

\newcommand{\bsub}{\begin{subequations}}
\newcommand{\esub}{\end{subequations}}

\DeclareMathOperator{\Tr}{Tr}

\def\vacS{|\textrm{vac}\rangle_S}

\begin{document}


\title{The maximum advantage of quantum illumination}

\author{Shannon Ray}
\affiliation{Information Directorate, Air Force Research Laboratory, Rome, NY 13441, USA}
\author{James Schneeloch}
\affiliation{Information Directorate, Air Force Research Laboratory, Rome, NY 13441, USA}
\author{Christopher C. Tison}
\affiliation{Information Directorate, Air Force Research Laboratory, Rome, NY 13441, USA}
\author{Paul M.  Alsing}
\affiliation{Information Directorate, Air Force Research Laboratory, Rome, NY 13441, USA}

\date{\today}

\begin{abstract}
\noindent Discriminating between quantum states is a fundamental problem in quantum information protocols. The optimum approach saturates the Helstrom bound, which quantifies the unavoidable error probability of mistaking one state for another.  Computing the error probability directly requires complete knowledge and diagonalization of the density matrices describing these states. Both of these fundamental requirements become impractically difficult to obtain as the dimension of the states grow large.  In this article, we analyze quantum illumination as a quantum channel discrimination protocol and circumvent these issues by using the normalized Hilbert-Schmidt inner product as a measure of distinguishability.  Using this measure, we show that the greatest advantage gained by quantum illumination over conventional illumination occurs when one uses a Bell state.
\end{abstract}

\maketitle

\section{Introduction}\label{sec:intro}
\noindent One of the main limitations to sending classical information using quantum states is the receiver's ability to distinguish the states carrying said information.  If these states do not have orthogonal support, there is an unavoidable probability that the receiver will mistake one state for another; this creates error in the message.  Therefore, it is necessary to have a measure that quantifies the probability of making an error, or a measure of distinguishability when analyzing which states are optimal for sending information.  

In 1969, Helstrom's work~\cite{helstrom} on the problem of discriminating between states $\Phi_0$ and $\Phi_1$ that are respectively sent with probabilities $p_0$ and $p_1$ established the Helstrom bound  
\begin{equation}
\label{eq:helstrom}
\underset{\{\Pi_i\}}{\min}\ p_E=\frac{1}{2}\left(1-||p_0 \Phi_0 - p_1 \Phi_1||_1\right)
\end{equation}
as the standard for quantifying the unavoidable error of mistaking one state for another.  Indeed, Eq.~\ref{eq:helstrom} is the minimization of the error probability
\begin{equation}
\label{eq:error}
p_E=p_0 \Tr \left[\Phi_0 \Pi_1\right] +p_1 \Tr\left[\Phi_1 \Pi_0\right]
\end{equation}
with respect to a set of positive operator value measures (POVMs) $\{\Pi_i \geq 0,i=0,1\}$ where $\Pi_0=\hat{\mathbf{1}}-\Pi_1$ and $\hat{\mathbf{1}}$ is defined as the identity operator.  In Eq.~\ref{eq:helstrom}, the trace norm $||\bullet||_1$ is defined as
\begin{equation}
\label{eq:tracenorm}
||\rho||_1\equiv\Tr[\sqrt{\rho^{\dagger}\rho}]
\end{equation}
where $\rho$ is an arbitrary operator and $\rho^{\dagger}$ is its Hermitian transpose.  Because the Helstrom bound is the standard for quantifying unavoidable error, most quantum information protocols that have a distinguishing process need to compute the trace norm, which requires diagonalization in general.  This can be difficult to work with when conducting an analysis especially as the dimension of the state becomes large.  One such class of protocols that require diagonalization is quantum channel discrimination (QCD).  The focus of this article is on the optimization of a specific QCD protocol. 

In QCD, one sends an input state $\Phi^{\left(in\right)}$ through a quantum channel which performs one of two operations on the state given by $\{ \mathcal{E}_i, i=0,1\}$.  They then receive the output state $\Phi^{\left(out\right)}_i=\mathcal{E}_i\left(\Phi^{\left(in\right)}\right)$ which is used to determine which operator acted on $\Phi^{\left(in\right)}$.  Of course, some input states will work better than others depending on the distinguishability of $\Phi^{\left(out\right)}_0$ and $\Phi^{\left(out\right)}_1$.  Here, the probability of mistaking one operation for another is quantified by the Helstrom bound
\begin{equation}
\label{eq:chandisc}
p'_E=\underset{\{\Pi_i\}}{\min}\ p_E = \frac{1}{2}\left(1-||p_0 \mathcal{E}_0\left(\Phi^{\left(in\right)}\right) - p_1\mathcal{E}_1\left(\Phi^{\left(in\right)}\right)||_1\right)
\end{equation}
where it is assumed that an optimal measurement scheme is used.  In this context, QCD can be understood as the problem of finding the input state that minimizes Eq.~\ref{eq:chandisc} over the space of all $\Phi^{\left(in\right)}$.  Moreover, extending the space of input states to higher dimension (including joint entangled states $\Phi_{q}^{(in)}$) can further reduce the error probability~\cite{kitaev,sacchi}. If one partitions the joint system into a signal subsystem and an idler subsystem, where the signal subsystem is sent as a probe, and the idler system is held in a local memory, when the signal returns, a joint measurement can be made; this changes Eq.~\ref{eq:chandisc} to
\begin{equation}
\label{eq:entchandisc}
p'_E = \frac{1}{2}\left(1-||p_0 \left(\mathcal{E}_0\otimes \hat{\mathbf{1}}_I\right) \Phi^{\left(in\right)}_q - p_1\left(\mathcal{E}_1\otimes \hat{\mathbf{1}}_I\right) \Phi^{\left(in\right)}_q||_1\right)
\end{equation}
where $\hat{\mathbf{1}}_I$ is the identity operator on the idler subsystem. In this article, we analyze a post-selected model of quantum illumination (QI) as a QCD protocol where Eq.~\ref{eq:entchandisc} is minimized in the space of all $\Phi^{\left(in\right)}_q$.  

In Seth Lloyd's seminal paper \cite{lloyd:QI} on QI, the experimenter uses an biphoton $d$-mode Bell state to enhance the detection of a potential surface in a noisy background (See Fig.~1 for diagram).  Formulating the problem with the simplest possible mathematical treatment, Lloyd assumes that a single-photon is detected per trial if anything is detected at all.  This detection may be due to a returning signal or surrounding noise.  In our treatment of QI as a QCD problem, we denote the scenario of receiving a mixture between signal and noise by the operation $(\mathcal{E}_0\otimes \hat{\mathbf{1}}_I)\Phi_q$, and the operation where the surface is not present and only noise is detected as $(\mathcal{E}_1\otimes \hat{\mathbf{1}}_I)\Phi_q$.  Here, $\Phi_q$ is an arbitrary biphoton entangled state that is not necessarily a Bell state.    

To remove the restriction of single-photon detection, it was suggested that a full Gaussian-state analysis of QI should be conducted; such an analysis was completed by Tan et al.~\cite{gaussillum}.  Using $M$ copies of signal and idler beams obtained from continuous-wave spontaneous parametric down-conversion in the absence of pump depletion, they demonstrated an improvement in reducing the upper bound of the unavoidable error probability over a strictly coherent source.  Our analysis is restricted to the single photon discrete-variable setting where it is easier to develop arguments based solely on dimension and quality of entanglement without choosing a specific state. 
\begin{figure}[t]
\centering
\includegraphics[width=\columnwidth]{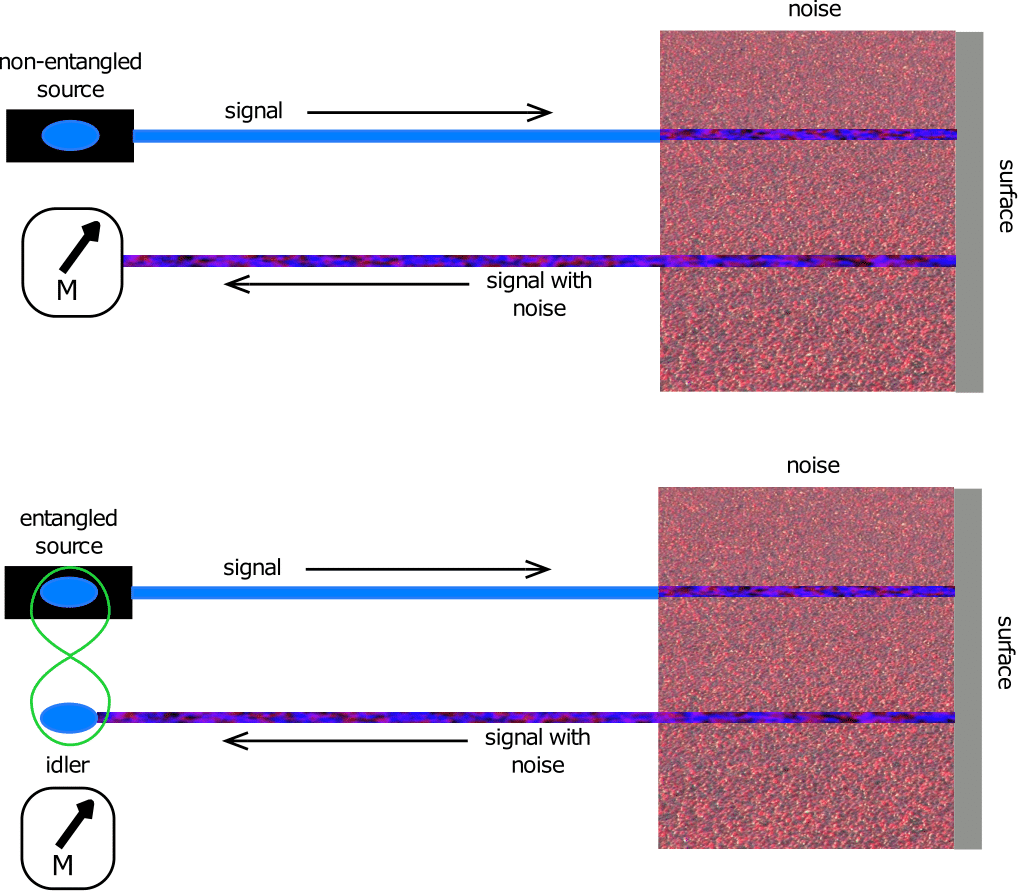}
\caption{(color online) (Top) Diagram of conventional illumination of target using conventional (separable states of) light. (Bottom) Diagram of quantum illumination of target with entangled states of light.  In the case of an entangled source, joint detection is performed between the held idler and returned noisy signal.}
\label{fig:tstarSqrd:vs:eta:tau}
\end{figure}
The relationship between our discrete-variable analysis to the continuous-variable setting will be a focus of future research.  

To avoid the problem of diagonalization when computing Eq.~\ref{eq:entchandisc} for the analysis of QI, we use the Hilbert-Schmidt inner product (HS), $\Tr[\rho^{\dagger} \sigma]$, to define a measure of distinguishability.  Since the HS inner product only requires the trace of a matrix product to compute, it significantly reduces the difficulty of analysis. One of the main goals of this paper is to demonstrate the efficacy of the HS inner product as a tool for discrimination. 

Given that the HS inner product significantly simplifies our analysis of QI (as we shall show), it may yet be used to simplify the analysis of other quantum information protocols.  In fact, this approach was used in~\cite{lee} as a measure of fidelity between a Bell state and its teleported counterpart, and it was used in~\cite{popescu} to avoid the trace norm when quantifying the average distance between two states.  Although the HS inner product satisfies Josza's axioms~\cite{fidelityref} of a fidelity measure, it does not increase monotonically under general quantum operations~\cite{ozawa,fidelityref}. This is important, where the action of a quantum channel on a pair of quantum states cannot increase their distinguishability (or decrease their fidelity).  Fortunately, for the class of states considered in the model of QI considered here, we show that the normalized HS inner product is monotonic with respect to its parameterization.

In this article, we analyze a modified model of Lloyd's original QI formulation.  Not only do we seek the states $\Phi^{\left(in\right)}_q$ that minimize Eq.~\ref{eq:entchandisc} for this model, we also show that the $d$-dimensional Bell state, defined as a maximally entangled state with equal-dimension subsystems, gives the greatest advantage of QI over conventional illumination (CI).  Conventional illumination uses the same input signal as the entangled case, but there are no idlers held to increase its effective brightness; the advantage is defined as the difference in distinguishability between signal and noise as given by QI versus CI.  

This article is structured in the following way.  In the next section, we present some background on QI and the mathematical framework used to conduct our analysis.  After that, we introduce the HS distinguishability measure and show that it reduces the analysis of QI as a QCD protocol entirely in terms of dimensional arguments and the purity of the ancilla/idler subsystem.  After that, we present the result that the $d$-dimensional Bell state gives the greatest advantage over CI for any other choice of $\Phi^{\left(in\right)}_q$.  This agrees with the recent results of De Palma and Borregaard~\cite{PB:ULTPREC} where they used asymmetric hypothesis testing~\cite{asymmetric} to show that the two-mode squeezed state gives the greatest advantage of QI.  These results are consistent since the two-mode squeezed state is the continuous-variable analogue to the $d$-dimensional Bell state.  Finally, we conclude with a discussion on the advantages of using a Hilbert-Schmidt based measure to address the problem of discrimination and its possible applications to quantum information protocols beyond QI.  

\section{Quantum Illumination}\label{sec:QI}
\noindent In this section, we describe the model used to analyze QI.  To do this, we present the original formulation of QI by Lloyd~\cite{lloyd:QI}.  Then we discuss our model which is the post-selected model used by Weedbrook et.\ al~\cite{discord}.

In Lloyd's original formulation of QI, only single photon events are considered.  In this setting, the signal consists of a single photon in $d_S$ possible modes.  It also assumes that the detector can distinguish between $d_S$ modes and its detection window is set to only detect a single photon per trial.  In this formulation, Lloyd chooses $\Phi_q$ to be the $d$-mode Bell state, which is given by $\Phi_q=|\phi_{Bell} \rangle \langle \phi_{Bell}|$ where $|\phi_{Bell}\rangle=d^{-1/2}\sum^{d}_{k=1}{|\mathbf{1}_k\rangle_S |\mathbf{1}_k\rangle_I}$.  Here, $|\mathbf{1}_k\rangle_S$ is the state with exactly one photon in the signal mode $k$ and no photons in the other $d_S-1$ modes.  A similar description is given for $|\mathbf{1}_k \rangle_I$.  Next we describe how noise is modeled in this setting.

When the signal is lost due to the target being absent, the remaining state is given by
\begin{equation}
\label{eq:remain}
\rho^{(1)}=(\mathcal{E}_1\otimes\hat{\mathbf{1}}_I)\Phi_q=\left((1-\lambda)\,|\textrm{vac}\rangle \langle \textrm{vac}| 
+ \lambda\,\frac{\hat{\mathbf{1}}}{d_S}\right)_S\otimes\Phi_I
\end{equation}
where $\lambda$ is the average number of noise photons received over many trials, $\vacS$ represents the vacuum where no photons are found in any of the signal modes, $\hat{\mathbf{1}}_S/d_S$ is the state representing the detection of a random mode from the surrounding noise, and $\Phi_I=\Tr_S[\Phi_q]$ is the idler subsystem held in local memory.  If the surface is present, the returning signal is a mixture between signal and noise, which is given by
\begin{equation}
\label{eq:returned}
\rho^{(0)}=({\mathcal{E}_0\otimes\hat{\mathbf{1}}_I})\Phi_q=\eta \Phi_q+(1-\eta)\rho^{(1)}
\end{equation}
where $\eta$, the average number of signal photons received over many trials, represents the degradation of the signal due to noise.   

In the paper by Weedbrook et.\ al~\cite{discord}, they use a simplified model of Lloyd's original formulation that assumes the detector always receives a photon either from the signal or surrounding noise.  They also complete their analysis using a Bell state, though they show that their result is independent of the state chosen in their appendix.  Their formulation of QI corresponds to a post selected model where $\lambda=1$ in Eq.~\ref{eq:remain}; this simplifies the remaining state to be 
\begin{equation}
\label{eq:onecon}
\rho^{(1)}=\frac{\hat{\mathbf{1}}_S}{d_S}\otimes \Phi_I.  
\end{equation}
Using this simplified model, they argue that quantum discord explains the underlying advantage of QI.  For computational clarity, we will also be working with this post-selected model.  

In the next section, we use the normalized HS inner product to show that the advantage of this post-selected model can be understood in terms of the purity of the idler subsystem given by $\text{Tr}[\Phi_{I}^{2}]$. In fact, a pure idler subsystem, (i.e., $\text{Tr}[\Phi_{I}^{2}]=1$) is necessarily uncorrelated from the signal, making the protocol using such states equivalent to CI.  Any value of $\text{Tr}[\Phi_{I}^{2}]<1$ implies an advantage gained by using a QI protocol. Where the minimum value of the purity for any density operator is $d^{-1}$, when $\Tr[\Phi^2_I]=d^{-1}_I$, the maximum advantage has been gained; this is equivalent to minimizing Eq.~\ref{eq:entchandisc}.  Unlike Weedbrook et.\ al and Lloyd, we do not assume $\Phi_q$ is the $d$-dimensional Bell.  Instead, we derive in Sec.~\ref{sec:proof} that this state gives the greatest advantage for this model.

\section{Hilbert-Schmidt Distinguishability Measure}\label{sec:prelims}
\noindent Between two arbitrary quantum states $\rho$ and $\sigma$, the normalized HS inner product is given by
\begin{equation}
\label{eq:normschmidta}
\langle \rho,\sigma \rangle \equiv \frac{\Tr[\rho^{\dagger} \sigma]}{\sqrt{\Tr[\rho^{\dagger}\rho]\Tr[\sigma^{\dagger}\sigma]}}.
\end{equation}
It has a lower extreme value of $0$ if and only if $\rho$ and $\sigma$ are states with orthogonal support \footnote{$\rho^{(0)}$ and $\rho^{(1)}$ having orthogonal support is equivalent to saying the probability that a system prepared in state $\rho^{(0)}$ will be measured to have an outcome associated to any of the eigenstates of $\rho^{(1)}$ is zero.}. It has an upper extreme value of unity if an only if $\rho$ and $\sigma$ are identical, and it is symmetric between them. The normalized HS inner product is invariant under unitary transformations, and it reduces to the ordinary inner product between quantum states when $\rho$ and $\sigma$ are pure. Moreover, we will show for the states $\rho^{\left(0\right)}$ from Eq.~\ref{eq:returned} and the remaining state $\rho^{\left(1\right)}$ from Eq.~\ref{eq:onecon}, that it is straightforwardly related to the physical parameters of QI.  Now we will write Eq.~\ref{eq:normschmidta} explicitly in terms of these physical parameters. 

To simplify Eq.~\ref{eq:normschmidta} and write it in terms of the physical parameters of QI, we replace $\rho$ and $\sigma$ with $\rho^{\left(0\right)}$ and $\rho^{\left(1\right)}$, respectively.  This is computed explicitly in the appendix.  Defining $\mathcal{H}_{01}$ as the normalized HS inner product between $\rho^{(0)}$ and $\rho^{(1)}$ to condense notation, our relations (from the appendix) simplify $\mathcal{H}_{01}$ to
\begin{equation}
\label{eq:distfinal}
\mathcal{H}_{01}\equiv\langle\rho^{(0)},\rho^{(1)}\rangle=\frac{1}{\sqrt{1+\eta^{2}\left(d_{S}K_{I}-1\right)}}
\end{equation}
where $K_{I}\equiv \Tr[\Phi_{I}^{2}]^{-1}$ is the inverse of the purity of the idler state $\Phi_{I}$.  Here, the physical parameters that completely characterize QI for a fixed $p_0$ are the relative signal fraction $\eta$, the dimension of the signal subsystem $d_{S}$, and the entanglement between signal and idler which is captured by $K_{I}$. Next, we want to show that both the minimum error probability $p_{E}'$ and $\mathcal{H}_{01}$ are extremized simultaneously with respect to these variables, so that we can use the distinguishability measure $\mathcal{H}_{01}$ to determine which states minimize the unavoidable error probability without diagonalization.  

To show that both $\mathcal{H}_{01}$ and $p'_{E}$ are extremized simultaneously, we must show that they are both monotonic with respect to parameters $\eta$, $d_S$, and $K_I$.  For a multi-variate function, we take montonicity to mean monotonic with respect to changes in each variable when all others are held constant.  One can verify that $\mathcal{H}_{01}$ is strictly monotonic by taking the gradient of Eq.~\ref{eq:distfinal} and showing that each term maintains the same sign over the intervals  $\eta \in [0,1]$, $d_S\in[2,\infty]$, and $K_I\in[1,d_I]$.  The interval for $d_S$ is justified if one assumes a qubit is the smallest signal used and one is allowed to use an arbitrary number of modes. 

From physical considerations, we can argue that given the values possible of the parameters, the error probability $p'_{E}$ monotonically decreases with increasing $\eta$, $d_{S}$, and $K_{I}$. Holding $d_S$ and $K_I$ fixed, it is clear that the error probability strictly decreases with increasing $\eta$ since it parameterizes the degradation of the signal due to noise.  As the signal becomes less noisy, it becomes easier to distinguish it from noise thus decreasing the chance of error.  Given that $d_S$ represents the possible modes the signal can be in as well as the number of modes distinguishable by the detector, holding $\eta$ and $K_I$ fixed only increases the dimension that is used to distinguish the known signal from surrounding noise.  Therefore, increasing $d_S$ strictly decreases the probability of mistaking the signal with noise.  Alternatively, lower-dimensional signals form a subset of higher dimensional signals, and expanding the set of states one is minimizing over cannot produce a worse result.  As in~\cite{popescu}, $K_I$ is the effective accessible dimension of the idler subsystem that expands the space of joint states obtainable through local manipulations of the signal subsystem (e.g., as in dense coding). Let $d'_S\equiv d_S K_I$ represent the effective dimension of the signal subsystem.  From here, we see that $d'_S \in [d_S,d]$.  When $d'_S=d_S$, one can only reduce $\mathcal{H}_{01}$ down to an amount limited by the dimension of signal subsystem; this is equivalent to a CI protocol.  When $d'_S=d$, one has access to the entire dimension of the idler subsystem to minimize $\mathcal{H}_{01}$.  As $K_I$ increases, the accessible dimension of the signal increases thus decreasing the probability of mistaking signal from noise.   

Where both $\mathcal{H}_{01}$ and $p'_{E}$ decrease monotonically with respect to $\eta$, $d_{S}$, and $K_{I}$, we can reach the minimum $\mathcal{H}_{01}$ and $p'_{E}$ along parametric curves of increasing $d_{S}$, $K_{I}$, and $\eta$. Along these trajectories, $\mathcal{H}_{01}$ is monotonic with respect to $p'_{E}$. Because of this,  the set of values of $\eta$, $d_{S}$, and $K_{I}$ that minimizes $\mathcal{H}_{01}$ also minimizes $p'_{E}$.  Therefore, one only needs to consider $\mathcal{H}_{01}$ when seeking to minimize Eq.~\ref{eq:entchandisc}.  

Looking at Eq.~\ref{eq:distfinal}, for a fixed $\eta$ and composite dimension $d$, it is clear that the minimum possible value of $\mathcal{H}_{01}$ is taken when $K_I=d_I$.  Therefore, the states that minimize Eq.~\ref{eq:entchandisc} are those whose idler subsystems have minimum purity (and therefore maximum entanglement with the signal).  This is equivalent to illumination protocols whose remaining states, $\rho^{\left(1\right)}$, are maximally mixed. Although all protocols for which $K_I=d_I$ minimize the error probability for a fixed dimension $d$, one must maximize $d_I$ to maximize the advantage of QI.  In the next section, we use the Schmidt decomposition to show that the $d$-dimensional Bell state is the only state that both has a remaining state that is maximally mixed and maximizes the idler dimension $d_I$. 

\section{Proof the Bell state gives the maximum advantage }\label{sec:proof}
\noindent In the previous section, we showed that the advantage of QI is quantified by $K_I$, and when $K_I=d_I$, one has gained the maximum advantage to distinguish $\rho^{\left(0\right)}$ from $\rho^{\left(1\right)}$ for fixed values of $\eta$ and $d$.  Therefore, if two states of equal dimension both have remaining states that are maximally mixed, they will have the same value of $\mathcal{H}_{01}$, but their advantages may be different.  Under this circumstance, the QI protocol with the greater value of $d_I$ will have a greater advantage. 

Given an arbitrary entangled pure state $\Phi_q=|\phi\rangle \langle \phi|$, its Schmidt decomposition is
\begin{align}
 \label{eq:schmidtDecomp}
|\phi\rangle = \sum_{m=1}^{r_{\text{min}}} \sqrt{\lambda_m} |s_m \rangle_S|i_m \rangle_I, \quad \sum_m \lambda_m = 1
\end{align}
where $r_{\text{min}}$ is the minimum rank between $\Phi_S$ and $\Phi_I$, $|s_m \rangle_S$ and $|i_m \rangle_I$ are orthonormal eigenbasis vectors for the signal and idler subspaces, respectively, and $\sqrt{\lambda_m}$ are the real non-negative Schmidt coefficients.  From here, we see that one must have $d_S>d_I$ or $d_S=d_I$ to get $K_I=d_I$.  Otherwise, its greatest value is restricted by the rank of the signal subsystem. 

Assuming maximum idler rank $K_I=d_I$, and the circumstances $d_S>d_I$ or $d_S=d_I$, the latter case achieves the largest possible effective signal dimension of $d'_S=d_S^2$ for a fixed signal, $d_S$.  Because the $d$-dimensional Bell state by definition is the only state with $d_S=d_I$ and $K_I=d_I$, it gives the greatest advantage of QI over CI or any other choice of $\Phi_q$.   Thus, by the analysis in this section we have found that the $d$-dimensional Bell state minimizes the error probability in the case of post-selection on the biphoton section of the composite Hilbert space.

\section{Discussion}\label{sec:Conclusion}
\noindent In this article, we treated QI as a QCD protocol to determine which states minimize the error probability and give the greatest advantage of QI.  Most approaches that address this problem require some diagonalization process such as when computing the trace norm or relative entropy. To avoid this problem we used the normalized HS inner product as a measure of distinguishability, which only requires the trace of the matrix product between density operators.   

Using this HS distinguishability measure, we identified three parameters in the post-selected model of QI $(\eta,d_S,K_I)$ that completely determine the distinguishability between $\rho^{\left(0\right)}$ and $\rho^{\left(1\right)}$.  The most important of these parameters is $K_I=\Tr[\Phi^2_I]^{-1}$ since it quantifies the advantage of QI over CI.  When $K_I=d_I$, one gains the maximum advantage afforded by QI, and when $K_I=1$, $\Phi_S$ and $\Phi_I$ share zero entanglement, which is equivalent to using a CI protocol.  

Although our analysis was on QI, we believe that the HS inner product may have applications to other quantum information protocols.  Similar analysis using the HS inner product may be possible for other protocols that use distributed entanglement among ancilla states to gain an advantage when sending or receiving information.  It is our intention to extend this research by considering such applications.

\section{acknowledgments}
\noindent The authors wish to graciously acknowledge the anonymous referee for providing very insightful comments and suggestions.  We would also like to thank the referee for a more direct derivation of terms in the normalized Hilbert-Schmidt inner products between $\rho^{\left(0\right)}$ and $\rho^{\left(1\right)}$; this derivation now appears in the Appendix.  SR and JS would like to acknowledge support for the National Research Council Research Associateship Program (NRC-RAP).
PMA, CCT, and JS would like to acknowledge support of this work from
Office of the Secretary of Defense (OSD) Applied Research for Advanced Science and Technology (ARAP)
Quantum Science and Engineering Program (QSEP) and the Defense Optical Channel Program (DOC-P).
Any opinions, findings and conclusions or recommendations
expressed in this material are those of the author(s) and do not
necessarily reflect the views of Air Force Research Laboratory.

\appendix*
\section{Simplifying the Hilbert-Schmidt inner product in terms of $\eta$, $d_S$, and $K_I$}
\label{sec:proofB}


\noindent We wish to compute $\mathcal{H}_{01}$ from \Eq{eq:distfinal} as
\be{H01} 
\mathcal{H}_{01} = \frac{ \Tr[\rho^{(0)} \rho^{(1)}] } {\sqrt{ \Tr[\rho^{(0) 2}] \, \Tr[\rho^{(1) 2}]} },
\ee
where
\be{rho0:rho1}
\rho^{(0)} = \eta\,\Phi_q + (1-\eta)\,\rho^{(1)}\quad  \hbox{and}\quad \rho^{(1)} = \frac{\hat{\mathbf{1}}_S}{d_S}\otimes\,\Phi_I.
\ee
We tackle each of the HS inner products in the numerator and denominator of \Eq{H01} one at a time.

First, it will be useful to compute the inner product $\Tr[\Phi \rho^{\left(1\right)}]$ directly.
\begin{align}
\Tr[\Phi \rho^{\left(1\right)}]&=\sum^{d_I}_{m=1}{\sum^{d_S}_{j=1}{_{I}\langle i_m|} _{S}\langle s_j|\left(\Phi \frac{\hat{\mathbf{1}}_S }{d_S} \otimes \Phi_I\right)|s_j\rangle_S |i_m\rangle_I       }\\
&= \sum^{d_I}_{m=1}{_I{\langle}i_m|\left(\sum^{d_S}_{j=1}{\frac{_S{\langle}s_j|\Phi| s_j\rangle_S}{d_S}}\right)\otimes\Phi_I |i_m\rangle_I} \\
\label{eq:2proof}
&=\sum^{d_I}_{m=1}{\frac{_I{\langle}i_m|\Phi^2_I|i_m\rangle_I}{d_S}}=\frac{\Tr[\Phi^2_I]}{d_S}
\end{align}
where $|s_m\rangle_S$ and $|i_m\rangle_I$ are orthonormal bases of the signal and idler subspace, respectively. 

Next we compute the numerator $\Tr[\rho^{(0)} \rho^{(1)}]$  which gives
\begin{align}
\Tr[\rho^{(0)} \rho^{(1)}]&=\eta \Tr[\Phi \rho^{\left(1\right)}] + \left(1-\eta\right)\Tr[\rho^{\left(1\right)2}] \\
&=\eta \frac{\Tr[\Phi^2_I]}{d_S}+\left(1-\eta\right)\frac{\Tr[\Phi^2_I]}{d_S} \\
\label{eq:3proof}
&=\frac{\Tr[\Phi^2_I]}{d_S}.
\end{align}
In the above we have used the result
\be{rho1:sqrd}
\Tr[\rho^{(1) 2}]=\Tr_S[\hat{\mathbf{1}}/d_S^2] \, \Tr[\Phi_I^2] = \Tr[\Phi^2_I]/d_S,
\ee
which is also needed in the denominator of \Eq{H01}.

From Eqs.~\ref{eq:2proof} and~\ref{eq:3proof}, we see that the inner product between $\Phi$ and $\rho^{\left(1\right)}$ is equal to the inner product between $\rho^{\left(1\right)}$ and itself; this implies that $\Tr[\rho^{\left(0\right)}\rho^{\left(1\right)}]=\Tr[\rho^{\left(1\right)2}]$.

Lastly, computing $\Tr[\rho^{\left(0\right)2}]$ gives
\begin{align}
&\Tr[\rho^{\left(0\right)2}]=\\
&=\eta^2\Tr[\Phi^2]+2\eta\left(1-\eta\right)\left(\Tr[\Phi \rho^{\left(1\right)}]\right)+\left(1-\eta^2\right)\Tr[\rho^{\left(1\right)2}] \\
\label{eq:4proof}
&=\eta^2 +\left(1-\eta^2\right)\frac{\Tr[\Phi_I^2]}{d_S}.
\end{align}
Inserting Eqs.~\ref{eq:3proof} and~\ref{eq:4proof} into Eq.~\ref{H01} and simplifying terms gives Eq.~\ref{eq:distfinal}.

\bibliography{MaxAdvantageQI}
\end{document}